# Le travail collaboratif dans le cadre d'un projet architectural

**Marie-France ANGO -OBIANG**
**Doctorante**
ango@loria.fr

**Laboratoire Lorrain de Recherche en informatique et ses Applications (LORIA)**

Campus Scientifique BP 329, 54506 Vandoeuvre-lès-Nancy France. Tél : +33 3 83 59 20 87
Fax : +33383278319

## Résumé :

L'analyse des pratiques et des tendances des utilisateurs lors de la recherche d'information sur Internet permet de mettre en évidence plusieurs points. La recherche d'information devient performante après connaissance de la typologie des différents systèmes de recherche. Cette typologie favorise l'adoption d'une méthodologie de recherche que l'on peut caractériser par les systèmes pull, agents intelligents, etc. Par ailleurs, l'importance de la structure du document électronique, correctement élaborée en amont, favorisera un taux de pertinence supérieur pour retrouver les informations.

Dans notre article, la problématique tourne autour de l'étude du comportement des utilisateurs en situation de recherche d'information, ainsi que la constitution d'un pôle de ressources documentaires dans un cadre d'un projet architectural. On constate que l'évolution des ressources documentaires est liée aux technologies de l'information.

**Mots-clés :** Travail collaboratif, besoin informationnel, Images architecturales.

# 1. Introduction

L'élaboration d'un projet architectural commence à la naissance du projet à travers l'expression d'un besoin et des intentions du maître d'ouvrage. Le maître d'ouvrage définit dans le programme les objectifs de l'opération et les besoins qu'il doit satisfaire ainsi que les contraintes et exigences de qualité sociale, urbanistique, architecturale, fonctionnelle, technique et économique, d'insertion dans le paysage et de la protection de l'environnement, relatives à la réalisation et à l'utilisation de l'ouvrage. Le Maître d'Ouvrage a besoin d'être guidées par le Maître d'Oeuvre et l'entrepreneur pour se poser les questions les plus pertinentes, pour formuler leurs besoins profonds et éviter de se limité à exprimer des besoins de surfaces ou directement sur des solutions (Figure 1).

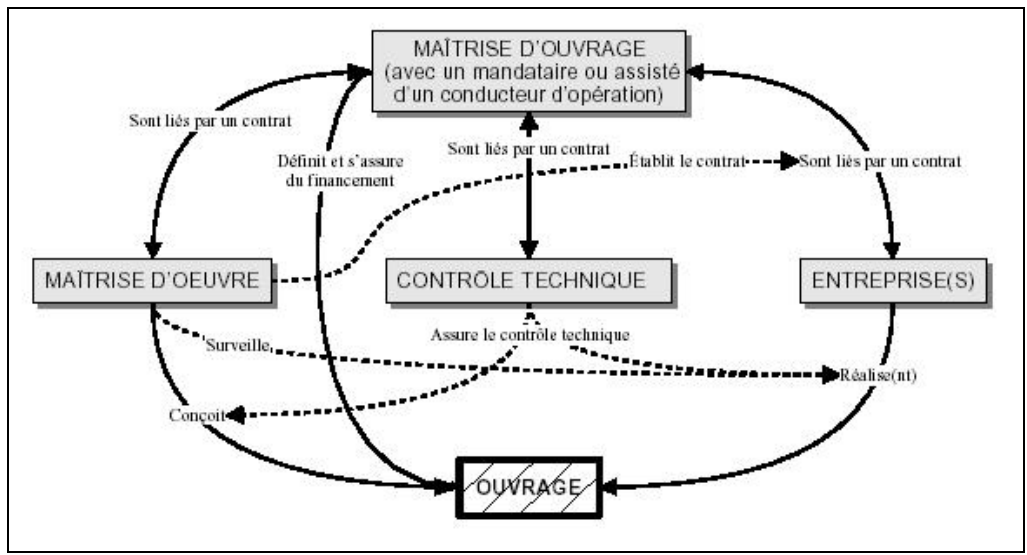

**Figure 1**: Rôle des acteurs entre eux et par rapport à l'ouvrage

Dès lors que le Maître d'Ouvrage a pu exprimer ces besoins réels, il a besoin d'être assistées pour les formaliser dans des cahiers des charges concis et précis. Il s'agit d'un point essentiel, car la réussite du projet et la coordination de tous les intervenants dépend beaucoup de la qualité des documents formels que sont les cahiers des charges. L'hypothèse de notre étude consiste à montrer que, si nous modélisons les besoins du Maître d'Ouvrage pour l'élaboration d'un projet architectural, nous aboutissons à une amélioration de la satisfaction des utilisateurs finals dans un système d'information qui s'applique dans le cadre du domaine architectural. Notre problématique est formulée selon la remarque suivante : Dans un système d'Information, dès lors que le Maître d'Ouvrage, le Maître d'Oeuvre et l'entrepreneur travaillent en collaboration, il est un rôle qui est prépondérant : le chef de Projet est le Maître d'Ouvrage. Nous pensons que aucun projet architectural ne peut être réalisé sans l'expression du besoin du Maître d'Ouvrage. De ce fait nous nous posons la question de savoir : comment représenter les besoins de nos utilisateurs dans un système d'information.

## 2. Les acteurs dans le domaine Architectural

La distinction entre Maître d'Ouvrage et Maître d'Oeuvre est essentielle dans le déroulement du projet, car elle permet de distinguer les responsabilités des deux entités. Il convient ainsi de s'assurer que l'expression des besoins reste sous l'entière responsabilité du Maître d'Ouvrage. En effet, il arrive dans certains cas que le Maître d'Ouvrage délègue au Maître d'Oeuvre des choix d'ordre fonctionnel sous prétexte d'une insuffisance de connaissances techniques, de façon concrète le service informatique d'une organisation prend la main et pilote le projet dès la phase d'expression des besoins.

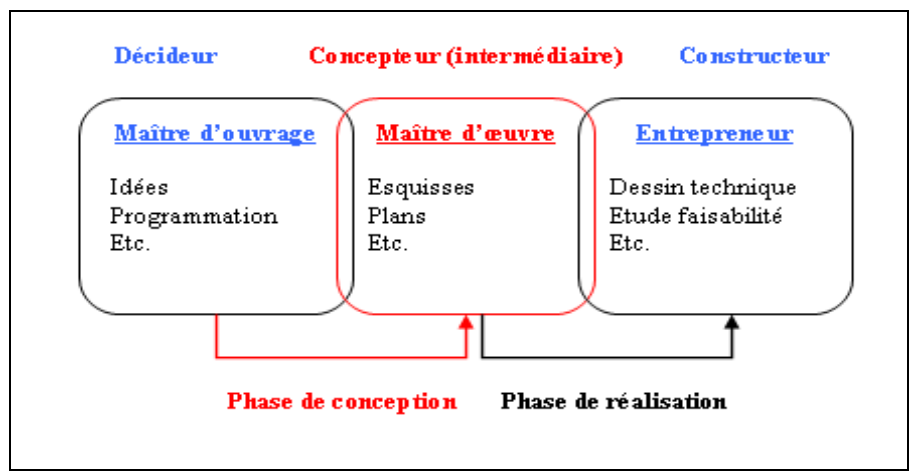

**Figure 2**: Maître d'Oeuvre rôle d'intermédiaire et de coordinateur entre le Maître d'Ouvrage et l'entrepreneur

Le Maître d'Oeuvre et l'entrepreneur sont les acteurs qui permettent de matérialiser le projet émis par le Maître d'Ouvrage (Figure 2). Cela n'est possible que si le Maître d'Ouvrage en fait la demande. Il est important de remarqué qu'aucun projet architectural ne peut-être élaboré, sans au préalable la mise en évidence des besoins et attentes du Maître d'Ouvrage. Le Maître d'Ouvrage est par conséquent l'acteur déclencheur d'action qui aboutit à l'élaboration d'un projet architectural.

## 2.1. Les besoins des acteurs

Répondre aux besoins exprimés le Maître d'Ouvrage c'est tout simplement lui fournir ce qu'il veut (la satisfaction du Maître d'Ouvrage) (Figure 3). En satisfaisant ses exigences formalisées dans un contrat ou un cahier des charges, le Maître d'Oeuvre adopte une démarche qualité. Les besoins implicites, (Bourdichon 1994) quant à eux sont tout simplement l'application des règles de l'art et des normes caractérisées dans la qualité d'un services rendus ; implicitement, lorsqu'un Maître d'Ouvrage veut l'acquisition d'une maison, la sélection d'effectue parmi différents entrepreneurs de bâtiment, etc. En fonction de son besoin, un usager se retourne naturellement vers celui qui est reconnu comme apte à la réaliser. Pour répondre aux besoins et attentes des usagers, les entreprises doivent, plus qu'à l'écoute, anticiper sur les évolutions du marché voire même les susciter par des opérations management judicieusement ciblées (Bouattour 2005).

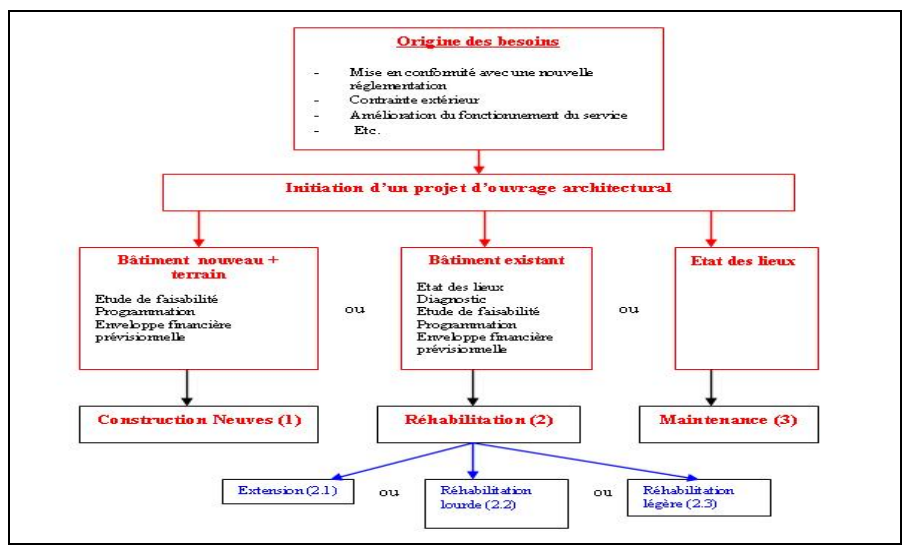

**Figure 3**: Organigramme des opérations d'un projet d'ouvrage architectural

Dans ce contexte, l'aspect communication prend toute sa mesure: l'information juste doit circuler rapidement au bon moment, à l'intérieur comme à l'extérieur de l'entreprise. Par ailleurs, il faut être sûr que l'acteur possède l'information correcte et s'assurer qu'après une transmission, l'information soit bien reçue et bien perçue. Cette importance accrue de l'information comme de sa circulation se trouve être encore renforcée par les besoins plus larges du marché. Les clients ne se suffisent plus de la seule possession du produit livré. Les exigences contractuellement exprimées dans la spécification de besoins résultent d'une consultation demandeur/concepteur, démarche consensuelle permettant d'obtenir le meilleur compromis entre le besoin réel, le coût et le délai.

## 2.2. Interaction entre les acteurs du bâtiment

Dans un schéma habituel, le Maître d'Ouvrage vient soumettre au Maître d'Oeuvre par exemple une intention d'habiter une maison. L'objet prend forme très progressivement et graphiquement par le savoir et savoir-faire de l'architecte. Les discussions et négociations avec le client (Maître d'Ouvrage) vont permettre à l'architecte ( Maître d'Oeuvre) d'affiner sa proposition en fonction de ce qu'il perçoit de la demande de son interlocuteur (Figure 4).

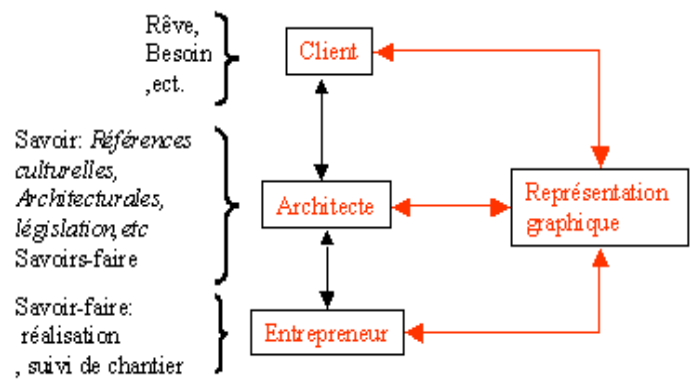

**Figure 4** : Interaction habituelle client/architecte/entrepreneur

Nous avons retenus trois acteurs (Maître d'Ouvrage, Maître d'œuvre, Entrepreneur) ([Ango-obiang 2006a) collaborant et intervenant par dialogue et/ou graphisme sur une même information visuelle qu'est l'image apporté par le client. Cette image du client est alors l'image référant de la consultation, image technique ou opérative (Lebahar 1983), il est aussi la trace visible et exploitable du passage de l'idée à la forme (dans la conception du client) ou de la forme à l'idée (dans l'explication qu'en fait le client à l'architecte). Il est surtout le support de la communication entre nos acteurs. Cette représentation sur laquelle les acteurs discutent et agissent n'a en fait aucune existence réalisée. En ce sens c'est un objet virtuel mais central, objet d'interaction où l'action est davantage orientée vers le partage, l'échange, l'interaction, la compréhension. En début de consultation, l'architecte se renseigne sur l'environnement du projet, le site, le terrain, l'orientation, le budget, etc.

Les réponses sont notées comme faisant parties des contraintes du projet. Très vite un véritable dialogue s'instaure entre les acteurs pour comprendre les logiques de chacun : l'architecte pour évaluer les «contraintes » de son client et le client pour saisir les modifications, rectifications qu'il doit apporter à son projet. L'architecte doit créer et rendre réalisable un édifice.

## 3. Le vecteur de la communication

L'acteur d'une activité constitue une image mentale de son environnement, des actions et des opérations à mettre en œuvre afin de réaliser son activité. Au cours de l'action, un acteur ne montre pas dans sa représentation mentale toute la complexité de l'objet ni toutes ses propriétés. L'acteur sélectionne l'information pertinente pour les actions qu'il souhaite mener sur l'objet. Au cours d'un travail collaboration, une orientation commune des sujets est nécessaire. Cependant, pour que l'accomplissement des actions puisse être coordonné, les orientations de tous les acteurs doivent être compatibles. Dans ce cas, le sujet doit former à la fois ses propres images opératives et se représenter les opérations, l'état des objets qu'il produit mais aussi les opérations et les objets produits par d'autres opérateurs. L'acteur doit donc échanger, communiquer avec les autres acteurs de l'activité. Les acteurs ne focalisent pas leur attention sur la communication, celle-ci demeure un outil au service de leurs actions. Au cours d'une activité la communication passe par la transmission de signes, qui, dans le cas d'une communication médiatisée par un langage sont porteurs de significations devant être communes à l'émetteur et au récepteur. La communication sert à la coordination en permettant l'échange d'idées ou de concepts à travers le dialogue et la transmission d'objets intermédiaires. (Michinov 2001) Les objets intermédiaires servent les activités de co-conception et de conception distribuée en assurant une synchronisation cognitive entre les acteurs impliqués.

## 3.1. Le concept objet intermédiaires

La communication au cours de la réalisation d'une activité collective nécessite également la transmission d'*objets intermédiaires* entre les acteurs. La conception d'un objet est « *ponctuée dans le temps* » par la production d'une quantité d'objets intermédiaires comme des idées, des textes, des dessins, des maquettes, etc. Ces objets intermédiaires sont « *des vecteurs de représentation, orientés par une intention ou un objectif issu d'un monde socio technico-économique lié d'une* façon *ou d'une autre à celui de la réalisation de cet objectif*» (Jeantet et al. 1996). Ces objets constituent donc la matérialisation des interactions apparaissant entre les acteurs au cours de la conception d'un objet. Les objets intermédiaires participent à l'orientation de l'activité en introduisant des interprétations, des matérialisations d'un état de l'activité en cours de réalisation.

## 3.2 Objets intermédiaires dans la communication

La conception est définie comme l'activité intellectuelle par laquelle sont imaginées quelques dispositions visant à changer une situation existante en une situation préférée (Simon 1991). En la contextualisant, elle peut être précisée de la façon suivante «*La conception consiste à donner un ensemble de propositions permettant de décrire le produit (forme, dimension, moyen d'obtention) et répondant globalement à un cahier des charges*» (Tichkiewitch al 1993). En effet, elle est située dans le monde des idées, de la connaissance, mais aussi elle relève aussi de la sphère de l'action. Ainsi ne faut-il pas la limiter à une activité intellectuelle : c'est en même temps une activité de création et de décision. En témoigne la production des multiples objets intermédiaires qui la compose, que ceux-ci soient immatériels (règlements, logiciels, modèles numériques,..) ou matériels (dessins techniques, textes, maquettes,..). Le but de notre présentation est de montrer que ces objets intermédiaires constituent les vecteurs les plus pertinents des activités de communication omniprésentes dans le processus de conception (Ango-Obiang 2006b). Dans la mouvance actuelle qui tend à intégrer de plus en plus ce processus, il devient urgent de prendre en compte la place qu'occupent les objets intermédiaires dans une communication devenue vitale pour la mise en place de la conception intégrée. Cet outil d'analyse doit nous permettre d'entrer au coeur de l'action de concevoir avec une double vue: la première orientée vers le contenu de la conception dont l'objet est la représentation, la seconde orientée vers les interactions entre les acteurs de la conception dont ces objets sont le centre. Ces objets intermédiaires (Mer et Tichkiewitch al. 1995) sont les «*vecteurs les plus pertinents des activités de communication omniprésentes dans le processus de conception* », ils sont à la fois ce qui va définir le produit lui-même et être le support de l'interaction entre les concepteurs ou partenaires.

## 4. Rôle de l'image

Dans le domaine du bâtiment, l'expression des contraintes et la recherche de solution passe prioritairement par des modes d'expression graphique. Pour les architectes, par exemple, les jeux permanents de références et d'analogies conduisent à penser que l'image est un objet intermédiaire qui occupe une place centrale dans leur stratégie de conception (Conan 1990), (Fernandez 2002). Ainsi, l'image permet de se confronter aux contraintes connues et de faire germer de nouvelles directions pour le projet. L'image est un moyen de communication très important dans le domaine architectural qui intervient dans les différentes phases de la conception d'un projet. Elle est considérée comme un outil d'aide à la décision. Les images sont des opérateurs puissants de conversion sensorielle. L'image est d'abord un objet auquel on se colle mentalement. C'est seulement dans un second temps qu'on s'en décolle en y appliquant des opérations de transformation. Et c'est alors qu'elle existe comme premier écran pour la pensée. On peut formuler les choses autrement. Les images que nous regardons n'ont pas seulement le pouvoir de donner du sens, mais ont aussi le pouvoir de nous contenir et d'exister comme lieu de transformations multiples.

## 4.1 L'image : support communicationnel de l'interaction

L'image est un support pour la recherche d'information lors de la phase de conception en architecture. Au cours d'une recherche d'informations techniques (idée, produit, exemple), l'utilisateur peut acquérir plus rapidement des informations avec des images d'œuvres d'architectures, qu'il ne le ferait à la lecture de textes (Nakapan et al 2002). De plus, l'image représente des informations parfois subjectives à caractère esthétique, qui ne peuvent pas être transcrites sous forme d'un texte. L'interprétation que l'on fait d'une image dépend de la culture et de la langue de la personne qui la regard. Toutefois, nous pouvons affirmer que « toute image est lisible quelque soit la langue ou la culture de la personne qui l'observe ». Dans toute tâche de conception, la visualisation est une nécessité absolue et concomitante.

Nous baserons notre propos sur l'image en tant que support de cette interaction. Principalement parce que l'image d'architecte est et reste la trace visible et exploitable du passage de l'idée à la forme ou de la forme à l'idée. Il est étroitement lié à une culture et à une technique de l'architecte. Présent à toutes les étapes de la conception, il est de fait le mode de communication par excellence, même si des explications souvent orales, parfois écrites peuvent l'accompagner.

## 4.2 L'usage de l'information véhiculée par les images

Nous assimilons une information à une collection de données organisés pour donner forme à un message, le plus souvent sous une forme visible, images, écrite, ou orale. La façon d'organiser les donnés résulte d'une intention de l'émetteur et est donc parfaitement subjective.

- *Une connaissance vient s'intégrer dans son système personnel de représentation, pour cela l'information reçue subit une série d'interprétation liées aux croyances générales, au milieu socioprofessionnel, au point de vue à l'intention, au projet de vue, a l'intention, au projet de l'individu récepteur. Pour qu'une information devienne connaissance, il faut également que le sujet puisse construite une représentation qui fasse sens*

- Contrairement *à l'information la connaissance n'est pas seulement mémoire, item figé dans un stock ; elle reste activable selon une finalité, une intention, un projet.*

Traiter l'information; c'est rassembler l'ensemble des données recueillies par les différents canaux pour en faire une synthèse, cohérente et surtout porteuse de sens pour l'utilisateur. Certains modèles d'évaluation des sources et de la valeur de l'information, ainsi que des outils d'aide à la sélection et à la lecture des systèmes d'informations, se sont développés au cours des dernières années. Ils présentent malheureusement le plus souvent des limites, spécifique à certains domaines d'activités.

## 5. Perspectives

Le besoin informationnel d'un utilisateur est un concept qui varie en définition, selon les chercheurs et selon les différents utilisateurs. Il existe des recherches qui ont essayé de lui donner une définition (Campbell et Rijsbergen 1996). Nous définissons un besoin informationnel comme étant une représentation informationnelle d'un problème décisionnel. Définir un problème décisionnel implique une connaissance sur l'utilisateur et son environnement. Un problème décisionnel peut être considéré comme une fonction d'un modèle de l'utilisateur, de son environnement et de son objectif. De ce fait nous proposons une méthodologie qui prenne en compte l'aspect multi-acteurs à la conception d'un projet.

### 5.1 Modélisations des l'acteurs

Nous évoluons dans le domaine architectural où il s'agit de mettre en relation l'information et les acteurs. Nous vous présentons comment la modélisation des besoins des acteurs (Maître d'Ouvrage, Maître d'Oeuvre, entrepreneur) permet d'améliorer les réponses. Nous avons mis en place le modèle SARBA (**S**ystème d'**A**ide de **R**eprésentation des **B**esoins informationnel en **A**rchitecture) qui se traduit par le fonction ; **RU = (T, B, F, A)** dans lequel est dressé une classification des acteurs où nous disons qu'un utilisateur (**U**) est représenté, par un type d'acteurs (**T**) {Maître d'ouvrage, Maître d'œuvre, Entrepreneur}, des besoins (**B**) {Projet de construction neuve, réhabilitation, extension, organisation, évaluation du budget}, la fonction (**F**) {Décideur celui qui dirige, Concepteur celui qui conseille, Développeur celui qui organise}, les activités des acteurs (**A**) {Annoter, intégrer, analyser, explorer, synthétiser} concernent leurs activités lors de l'utilisation du système d'information. Ils peuvent être amenés à rechercher, télécharger des dossiers, comprimer des fichiers, annoter des images ou du texte, indexer des documents, consulter des notes. Ils doivent parfois disposer d'un certain degré d'interactivité avec le système.

Les besoins des acteurs sont très variés allant, d'une simple découverte d'un sujet complètement nouveau, confirmation d'un savoir élémentaire, approfondissement de ce savoir à son niveau, exploration - découverte pour faire le point, étude de haut niveau. Les besoins par exemple d'un étudiant en architecture ne sont pas les mêmes que ceux d'un professionnel du bâtiment (Architecte, Entrepreneur). C'est dans ce sens que notre étude tant à comprendre les besoins spécifiques de nos acteurs dans la recherche d'information pertinente pour leur prise de décision. Cette dernière, ne peut être possible sans une analyse des différents besoins en information.

## 5.2. Modélisation des documents

Les ressources documentaires mises dans notre système d'information proposent différents types de documents à destination de types d'acteurs différents. On recense entre autres : des documents administratifs (dossier d'appel d'offre, permis de construire, etc.), des images (dessin technique, plan technique, esquisses, etc. .). Tous ces éléments sont au service de l'élaboration d'un projet architectural, pour les acteurs en situation de recherche d'information. L'objectif de la modélisation, puis de la description est de favoriser la visibilité d'un document tout en préservant une expression simple des informations pouvant répondre aux questions suivantes : quelles sont les caractéristiques de la ressource ? Comment classer cette ressource ? Comment mettre en relation le profil de la ressource et le profil utilisateur ?
Les documents sont déposés sur le Système d'Information par classification, avec indexation de documents, avec des information sur leur contenu, leur contenant, leur but (figure 5). Ces documents sont un corpus de d'images d'œuvres architecturales que nous avions pris dans des cabinet d'architectures, ministères de l'habitat et autres.

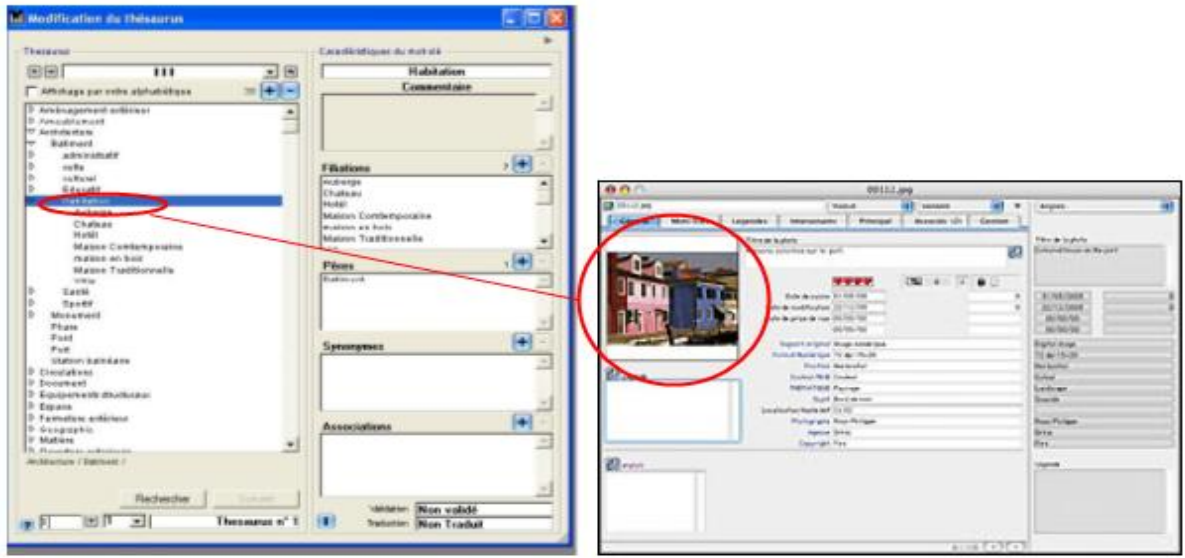

**Figure 5** : Indexation d'une image de bâtiment dans notre système d'information

## 5.3. Mise en relation du modèle des acteurs et du modèle des ressources documentaires

Notre domaine d'application (Ango-obiang 2004) met en situation un utilisateur, dont le profil est décrit dans des métadonnées, obtient des informations personnalisées. Ces informations peuvent concerner des ressources documentaires ou des informations techniques et financières. En fonction de son rôle, l'utilisateur pourra visualiser des niveaux d'information. Nous mettons en évidence que nos acteurs (maître d'ouvrage et maître d'œuvre) ont des besoins, des fonctions et des activités spécifiques.

L'étude des normes en vigueur des ressources documentaires, permet d'élaborer des classes d'objet. Nous répartissons ces classes d'objets autour de quatre pôles qui prennent en compte le contexte de l'utilisateur :

− **Propriétés de la ressource** : format, forme, identifiant.

− **Identification des besoins des utilisateurs auxquels répond un document** : ressource, objet à définir, public concerné.

− **Données de description relatives au contexte** : titre, mots-clés, thématique, légende, annotation.

− **Ressources documentaires autour d'un module** : activités, experts, densité sémantique.

Actuellement notre modèle a été inséré dans un outil informatique afin d'être expérimenté, par des utilisateurs qui sont des professionnelles du bâtiment.

## 6. Conclusion

Par ce papier, nous avons montré comment la modélisation des différents acteurs permettent d'élaborer les méta-données des acteurs pour intégrer leur type, leurs besoins, leurs fonctions et leurs activités. Nous avons montré comment il est possible de représenter les utilisateurs pour la mise en place des bases données métiers qui aboutir à un modèle formel, nommé SARBA. Nous mettons en évidence que ce modèle SARBA, permet de donner des vues différentes du SIS aux différents acteurs. L'aspect dynamique du modèle est dû au fait que le contenu de notre SI est guidé par les besoins des utilisateurs. Cette étape a favorisé la mise en relation des acteurs et des ressources documentaires.

## Bibliographie